# 4H-SiC Schottky diode radiation hardness assessment by IBIC microscopy

[1]Ettore Vittone, [1]Paolo Olivero, [2]Milko Jaksic, [3]Zeljko Pastuovic

[1]Physics Department, University of Torino (I), via P. Giuria 1, 10125 Torino, Italy.

[2]Laboratory for Ion Beam Interactions, Ruđer Bosǩović Institute, Bijenička 54, Zagreb, 10000, Croatia.

[3]Centre for Accelerator Science, Australian Nuclear Science and Technology Organisation, Lucas Heights, 2234, NSW, Australia

## Abstract

We report findings on the Ion Beam Induced Charge (IBIC) characterization of a 4H-SiC Schottky barrier diode (SBD), in terms of the modification of the Charge Collection Efficiency (CCE) distribution induced by 20 MeV C ions irradiations with fluences ranging from 20 to 200 ions/$\mu m^2$.

The lateral IBIC microscopy with 4 MeV protons over the SBD cross section, carried out on the pristine diode evidenced the widening of the depletion layer extension as function of the applied bias and allowed the measurement of the minority carrier diffusion lengths.

After the irradiation with C ions, lateral IBIC showed a significant modification of the CCE distribution, with a progressive shrinkage of the depletion layer as the fluence of the damaging C ions increases.

A simple electrostatic model ruled out that the shrinkage is due to the implanted charge and ascribed the perturbation of the electrostatic landscape to radiation-induced defects with positive charge state.

## Introduction

Silicon carbide (SiC) has highly desirable physical and chemical properties that make it an ideal material for use in a wide variety of applications, ranging from micro-electromechanical systems [1], chemical sensing [2], quantum [3] and electronic devices. In this last field, silicon carbide, and in particular the 4H-SiC allotrope, plays a key role in power electronics [4] due to its large indirect band gap (~3.2 eV), large breakdown electric field (2 MV cm$^{-1}$), high electron mobility (900 cm$^2$ V$^{-1}$ s$^{-1}$) and high thermal conductivity (400 W m$^{-1}$ K$^{-1}$) [5]. One of the main elements of interest in this material is represented by the possibility of making SiC-based systems operating at high temperature [6] and, in general, in harsh environments [7], including the development of detectors operating in harsh ionizing radiation conditions [8]. Thanks to its high threshold energy for atomic displacement (24 eV for C and 66 eV for Si [9], to be compared with 21 eV in Si [10] and 50 eV for diamond [11]), SiC is an attractive candidate for the fabrication of radiation detectors [12] for applications ranging from laser-generated plasma diagnostics [13], high energy physics experiments [14], betavoltaics [15], x /gamma ray spectroscopy [16] to neutron monitoring [17] in fission reactors, in high-level radioactive waste [18] and fusion reactors [19], [20].

The assessment of SiC radiation hardness is therefore important for the development of SiC detectors in radiation-intensive environments, and has been the object of several studies, aimed to identify main defects and the relevant recombination levels induced by radiation, which degrade the electronic properties of SiC devices [21] , [22] , [23].

This paper contributes to this subject by reporting the characterization of a 4H-SiC Schottky diode by the Ion Beam Induced Charge technique (IBIC). The advantages of this technique stems from using MeV ions both as damaging and probing agents, which allow both the characterization of the transport and electrostatic properties of semiconductors, as well as the study of damage induced by ion irradiation [24], on the availability of a robust interpretative model [25], and on the possibility of quantitatively mapping the charge collection efficiency (CCE) by using scanning focused ion beams [26].

The experiment here presented and discussed, consists in the IBIC analysis of a 4H-SiC Schottky diode, which was cleaved in order to expose its lateral cross section to 4 MeV proton microbeam to probe the CCE profile (first phase). Subsequently, selected areas of the frontal

Schottky electrode were irradiated with 20 MeV C ion beam with different fluences (second phase). Finally, lateral IBIC analysis was performed in order to investigate the effects of the C ion irradiations on the CCE profiles (third phase).

## Experimental

The device under study was a a 4H-SiC Schottky diode fabricated on an n-type (~50 μm thick) epitaxial layer on low micropipe density (16–30 $cm^{-3}$), $n^+$ substrate from CREE Research Inc.. The Schottky and ohmic electrodes (3 mm in diameter) were deposited by evaporating Ni (100 nm) amd Au (100 nm) on the epitaxial layer and a Ti (30 nm) /Pt (30 nm) /Au (150 nm) on the substrate side (C-face), respectively (Fig. 1a,b). An almost constant effective donor concentration of about $1.5 \cdot 10^{14} cm^{-3}$ in the first 30 μm of the n-type region, an ideality factor of (1.14±0.03) and a barrier height of (1.52±0.01) eV were estimated from capacitance and current-voltage characterization.

The diode was cleaved in order to expose its lateral cross section to the ion beam irradiation (lateral IBIC set-up [27]). The cleavage increased the current in the reverse bias conditions, as shown in Fig. 1c, but at 100 V the current was maintained below 10 nA, i.e. values still compatible with IBIC measurements (Fig. 1).

The experiments were carried out at the RBI (Ruđer Bošković Institute) ion microprobe facility. The electronic chain to process the IBIC signal was constituted by a charge sensitive preamplifier (Canberra 2004) and a shaping amplifier (Canberra 2021, shaping time 1.5 μs). The Spector [28] hardware and software system was used for data acquisition and beam scan control. The calibration of the electronic chain was performed by using a reference Si detector, resulting in a charge sensitivity of 1800 electrons /channel and a spectral resolution (FWHM) of $1.2 \times 10^4$ electrons, which correspond to 14 keV/channel and 94 keV, respectively, in SiC, assuming an average energy to create an electron/hole pair of 7.78 eV [29].

The 4H-SiC SBD diode was irradiated at normal incidence through the surface Schottky contact with 20 MeV carbon ions, with the purpose of creating displacement damage in the active region under reverse biasing. Lateral IBIC measurements to characterise changes in CCE and electric field profile across the depletion region of the SBD were carried out using the probing 4 MeV proton microbeam focused to a spot size of about 2 µm (FWHM), as estimated from of an on-axis scanning transmission ion microscope (STIM) image of a copper grid (2000 mesh pitch). Since the electron/hole generation occurs primarily at the Bragg peak, i.e. at the end of the proton range (about 100 µm from SRIM simulations[30]), surface recombination at the irradiated (cleaved) surface is assumed not to play a dominant role.

## Results and Discussion

### First phase: IBIC analysis of the pristine diode

The characterization of the 4H-SiC diode by irradiating the cleaved sample from the side (lateral IBIC) was performed using a focused 4 MeV proton beam. The CCE maps (median values) at different bias voltages ($V_b$) shown in Fig. 2 evidence the presence of a region with high CCE, which widens as $V_b$ increases, followed by a rapid decrease in efficiency, as the distance (y) from the Schottky electrode increases. The CCE profiles, calculated along the y axis of the IBIC maps, corroborate these observations, as shown in Fig. 3.

The interpretation of these profiles can be first approached by the drift-diffusion model, as already adopted in previous IBIC analyses of Si p-n junction and GaAs Schottky diodes [31]: the plateau at 100% efficiency cover the depletion region, where the strong electric field very efficiently drifts both the minority (hole) and majority (electron) charge carriers generated by ionization towards frontal (holes) and back (electrons) electrodes, inducing a charge signal at the sensitive (frontal) electrode, which is processed by the electronic chain. The CCE decay at longer distance from the Schottky barrier is to be attributed to the probability of minority carriers, generated in the neutral region, to diffuse to the edge of the depletion region [32]. The logarithmic slope of these CCE decay curves is almost constant for all bias voltage values and is equal to the inverse of the hole diffusion length $L_h$, which can be estimated as $L_h=(4.9\pm0.3)$ μm.

More elaborated simulations, based on the theory IBIC theory derived from the Shockley-Ramo-Gunn theorem, have been carried out by solving the coupled Poisson's, the semiconductor continuity equations and the relevant adjoint equations [33]. These equations have been numerically solved by the finite element method [34], taking as input the doping profile and the Schottky barrier potential value extracted from capacitance–voltage measurements and the hole carrier lifetime ($\tau_h = 80$ ns). The latter value was derived from the diffusion length previously calculated, assuming a diffusivity of 3 cm$^2$/s [8]. The results of these 1D simulations, shown as continuous lines of Fig. 3a, are in good agreement with the experimental data. An alternative analysis of this device was carried out by J. Forneris et al. [35] by a Monte Carlo method.

**Second phase: Frontal electrode irradiation**

The irradiation of the frontal electrode was carried out by using 20 MeV C focused ion beam, in order to induce radiation damage in selected regions at different fluences ranging from 20 to 200 ions/μm$^2$.

Since the ion current was of the order of 10$^3$ ions/s, the IBIC signals were recorded during the irradiation, both to monitor the CCE degradation and to refine the fluence measurements, which were estimated by the number of detected ions divided by the area of the selected region (namely, 145×68 μm$^2$). Fig. 4a) shows an example of IBIC spectra taken during the irradiation of region E; the degradation of the median CCE monitored during the irradiation of the 4 regions is shown in Fig. 4b).

After the irradiation, IBIC maps of the sample were taken at different reverse bias voltages (namely, 10 and 50 V) using the same ion beam (20 MeV C), as shown in Fig. 5.

As expected from previous studies on the radiation hardness of silicon diodes [24], it is apparent that the smaller is the applied bias voltage, the higher is the contrast of the irradiated region in the IBIC map. The depletion layer at a reverse bias voltage of 10 V has a width of ~10 µm (see Fig. 3b). Therefore, a non-negligible fraction of charge carriers is generated by ionization in the neutral region, and in particular at the peak of the vacancy profile (Fig. 6), i.e. in proximity of the highest concentration of damage-induced defects. In this region, the slow diffusion of holes toward the depletion region fosters the recombination, the reduction of the hole lifetime and thus the degradation of the CCE. At a reverse bias voltage of 50 V, the depletion layer extends to ~20 µm (see Fig. 3b) and the generation profile is fully included within the depletion region, where the strong electric field promotes a fast drift of the carriers, which makes the recombination effective only in the regions irradiated at the highest fluences.

**3$^{nd}$ phase: Lateral IBIC of the irradiated regions**

In order to gain insight on the origin of the CCE degradation observed during the second phase of the experiment, lateral IBIC analysis of the irradiated regions was carried out, replicating the conditions adopted in the first phase. Fig. 7 shows the CCE maps obtained in lateral IBIC geometry, using 4 MeV protons as the ion probe, at different bias voltages of regions C,D,E,F, that were previously irradiated with 20 MeV C fluences of 20,61,180,200 ions/µm$^2$, respectively. It is apparent that the irradiated regions show a degradation of the CCE profile, which is more evident for higher fluences.

However, it is worth noting that the depth at which full induced charge collection is measured shrinks as the fluence increases, down to a depth (vertical coordinate in Fig. 7) which corresponds to the range of 20 MeV C ions (12.8 µm) in the diode. At the highest fluence (region F), the CCE is higher at this depth, where the vacancy peak, and then the highest recombination, occurs.

The CCE profiles, extracted from the central parts of the irradiated regions, confirm these observations, as shown in Fig. 8.

Looking specifically at the regions C and D (fluence 20 and 61 ions/μm², respectively) for $V_b$=50 and 100 V, the CCE exponential decay occurs at a depth which is larger than the C ion range (R). There, the lifetime values (and thus diffusion length values) of the minority carriers are not affected by damage effects and the logarithmic slopes of the CCE profiles in these two regions at the highest bias voltages are almost identical.

It is apparent that the shrinkage of the depletion region is the dominant effect of the 20 MeV C ion irradiation, and it can be ascribed to a modification of the diode electrostatics landscape- Actually, this evidence can be interpreted by a simple electrostatic model, which assumes that the modification of the effective charge density is induced by the damage proportional to the ion fluence confined in a planar interface locate at peak of the damage production (Fig. 6). This assumption can be considered a good approximation to the real damage profile, since the longitudinal straggling of the damage profile is small in comparison to the projected range [30]. Assuming a surface charge located at the C ion range (R=12.8 μm) emerging from a flat donor distribution ($N_D=1.5\cdot10^{14}$cm⁻³), the shrinkage can be related to the ion fluence through the expression

1) $(w_0^2 - w^2) = \frac{2R}{N_D}\alpha\varphi$

where α it the ratio of the surface density of positive charge generated by ion damage and the C ion fluence ($\varphi$). The linear fit shown in Fig. 8d provides a value of α of about 70, a value which rules out the hypothesis that the shrinkage of the depletion region is merely due to the implanted charge. Therefore, these positive charges could consist of radiation-induced electrically active defects, whose density is related to the number of defects generated by single 20 MeV C (Fig. 6), given the number of displaced atoms per ion (about 200 Si and 400 C), as computed by SRIM simulation [30]. Since such defects increase the space charge in the depletion region, we infer their positive charge state.

The change in effective doping in the space-charge region induced by radiation induced defects is a well-known effect in semiconductor detectors [36] and direct evidence of local effects of electrically active defects on the internal electric field has been observed in diamond [37]. However, our interpretation is in apparent disagreement with previous studies, which ascribe the irradiation effects in Schottky [22] and p-n [23] 4H-SiC diodes, irradiated with 7 MeV C and 700 keV He and ions, respectively, to doping compensation. However, the studies of Izzo et al. [22] and Pellegrini et al. [23] are based on the analysis of the current-voltage characteristics in forward steady-state conditions, and the apparent increase of the series resistance is associated to a decrease in the concentration of free carriers in the neutral region. It is worth remarking that the IBIC analysis described in this paper was performed in reverse polarization condition and the shrinking of the region where CCE=100%, i.e. the region where carriers induce charge at the sensing electrode, suggests a modification of the electrostatics, which does not necessarily coincide with a modification of the free carrier concentration.

Our model can be extended to interpret all the other profiles shown in Fig. 8. At $V_b$=20 V, the logarithmic slope of the profile in region C is apparently steeper than that the one which is relevant to the pristine region. Taking into account that this exponential decay occurs in the neutral region, where diffusion mechanism inject the minority carriers in the depletion region, the more rapid decay of the CCE profile can be ascribed to the increase of the recombination centers generated at the end of the ion range, which decreases the minority carrier lifetime and diffusion length. As the C ion fluence increases (regions D, E, F), the progressive decrease of lifetime value degrades the drift length of both the carriers and then the IBIC signal, as induced by the drift of charge carriers in the depletion regions. Similar consideration can be applied also for $V_b$=50 V and $V_b$=100 V. In particular, it is worth noting that in region E, the CCE profile is peaked at the depth corresponding to the C ion range, where the recombination should be maximum. These counter-intuitive results can be explained by assuming that, consistently with our model, at this depth a significant increase of the electric field occurs, which compensates the reduction of drift length due to the degradation of the carrier lifetime.

## Conclusions

In this work, IBIC analysis of a 4H-SiC Schottky diodes was performed in order to investigate the performances of this device in view of its application as ionizing particle detector. The experiments were conducted in lateral geometry, i.e. by raster scanning the 4 MeV proton microbeam over the cleaved cross section of the diode.

The CCE maps of the pristine diode show regions of maximum efficiency which widens as the reverse applied bias increases, in agreement with the results emerging from the standard electronic characterization; the analysis of the CCE exponential decay in the neutral region allowed the calculation of the minority carrier diffusion length.

In order to evaluate the effects of high energy ion irradiation on the device performances, selected areas of the diode were exposed to 20 MeV C ion beam at different fluences. The following lateral IBIC experiments showed a significant modification of the CCE distribution which could not be interpreted by ascribing the CCE degradation only to the increase of recombination centres induced by the radiation. Actually, the most evident effect at low fluence and high bias voltage was represented by the shrinkage of the depletion layer width, which suggests a modification of the diode electrostatics landscape induced by defects generated by the 20 MeV C ion irradiation. A simple model was introduced, which describes the electrostatic effects assuming a planar interface placed at the depth of maximum damage with a charge surface density proportional to the ion fluence. The results of this analysis indicate that this shrinkage is due to the charge implanted by the C ions and can reasonably be ascribed to the creation of defects assuming positive charge states in the space charge region. Such an interpretation is further corroborated by the presence of a peak of CCE located at the 2 MeV C ion range: the increase of the electric field in the space charge region compensates the reduction of the carrier lifetime, which was expected to degrade the carrier mean drift length and then the induced charge signal.

These results evidence radiation induced effects not considered in previous studies, and can trigger further analyses to assess the radiation hardness of 4H-SiC devices.

## Acknowledgement

This work was supported by the project 'Departments of Excellence' (L. 232/2016), funded by the Italian Ministry of Education, University and Research (MIUR). We would like to acknowledge the Australian Government funding for ANSTO Centre for Accelerator Science through the National Collaborative Research Infrastructure Strategy (NCRIS) project.

## List of references

# Figures

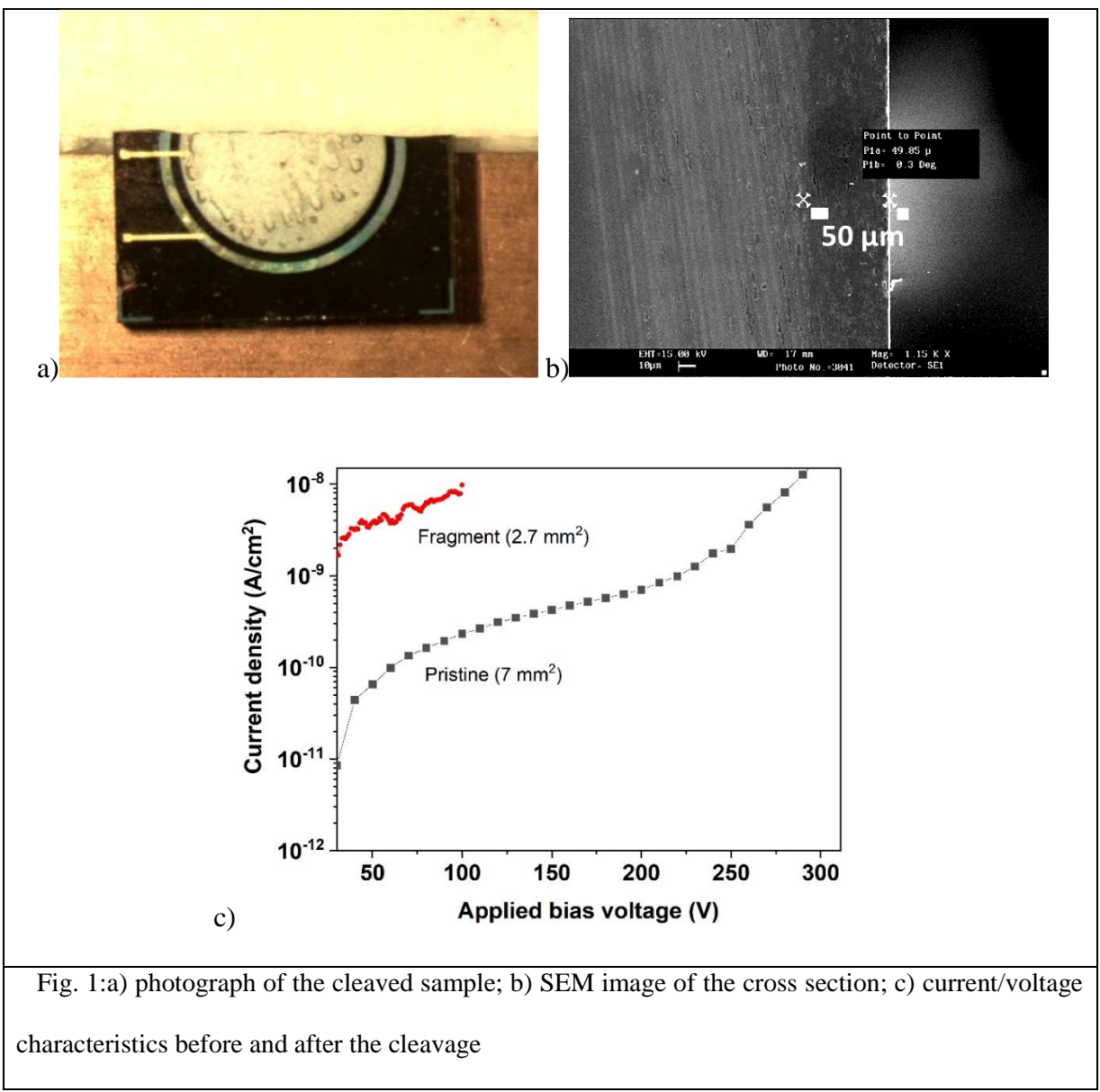

Fig. 1:a) photograph of the cleaved sample; b) SEM image of the cross section; c) current/voltage characteristics before and after the cleavage

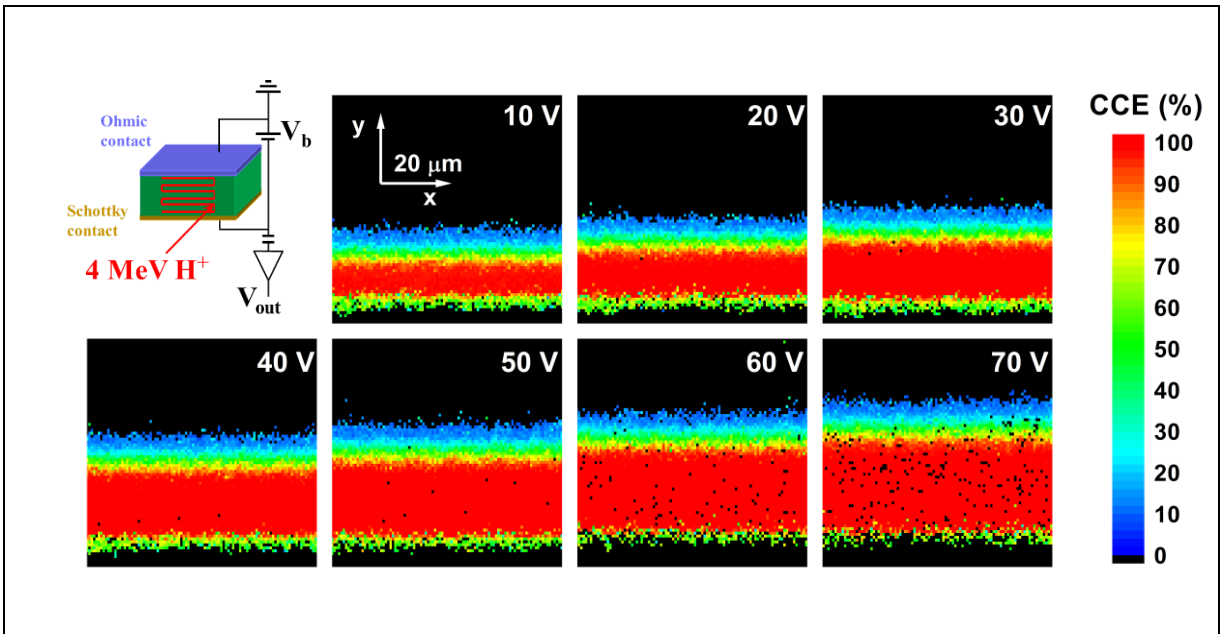

Fig. 2: : Scheme of the lateral IBIC set-up (top left) and CCE distributions resulting from lateral IBIC experiment at different applied bias voltages ($V_b$). The Schottky electrode is at the bottom of the coloured maps.

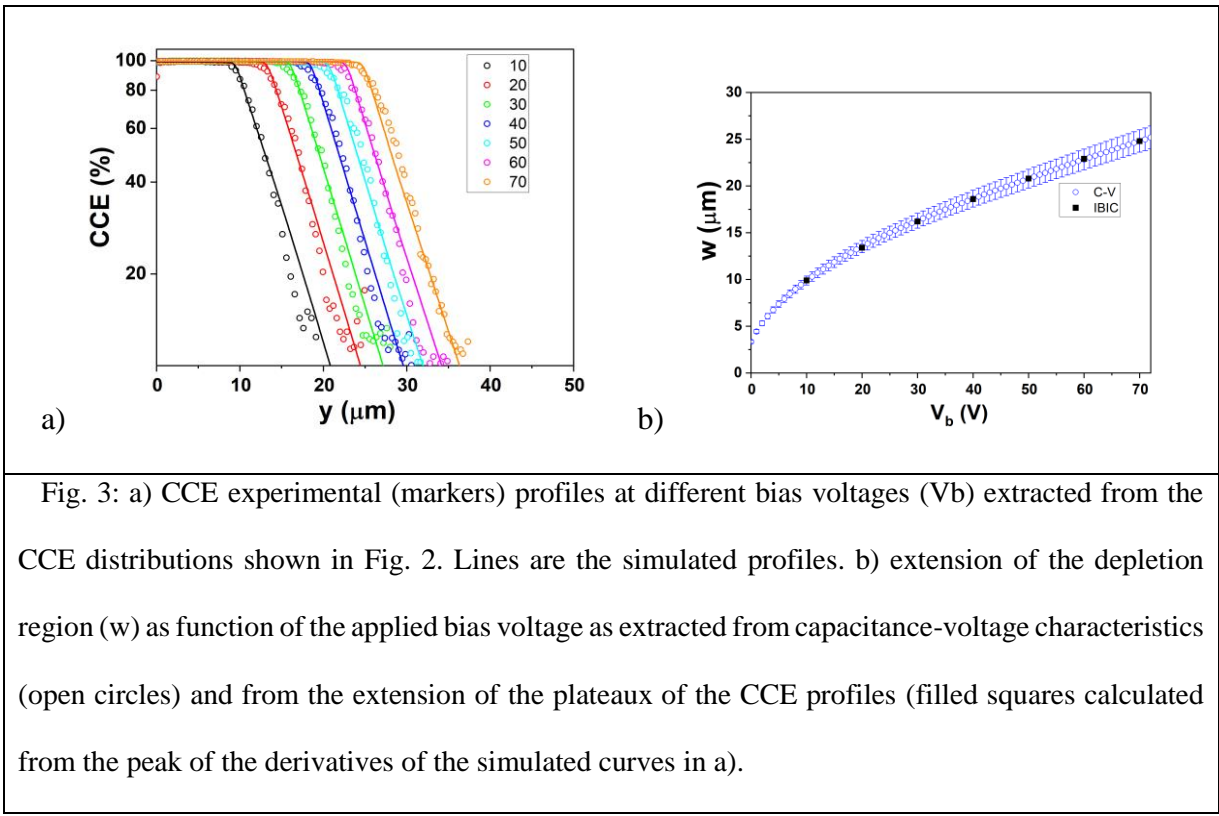

Fig. 3: a) CCE experimental (markers) profiles at different bias voltages (Vb) extracted from the CCE distributions shown in Fig. 2. Lines are the simulated profiles. b) extension of the depletion region (w) as function of the applied bias voltage as extracted from capacitance-voltage characteristics (open circles) and from the extension of the plateaux of the CCE profiles (filled squares calculated from the peak of the derivatives of the simulated curves in a).

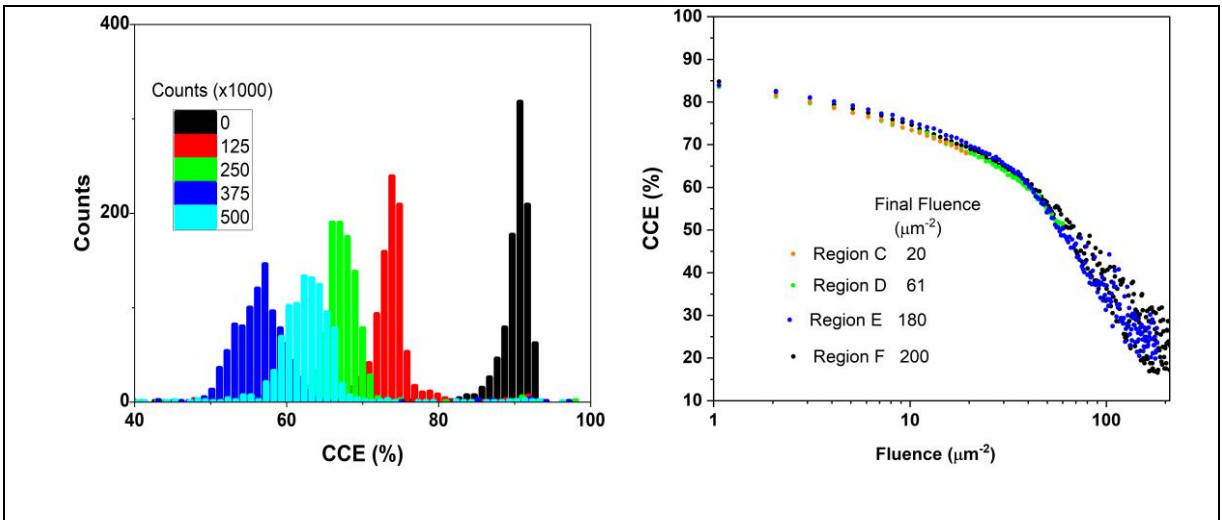

Fig. 4: a) IBIC spectra (1000 counts) taken during the irradiation of the region E at different fluences. b) evolution of the median CCE as function of the fluence of the six different irradiated regions

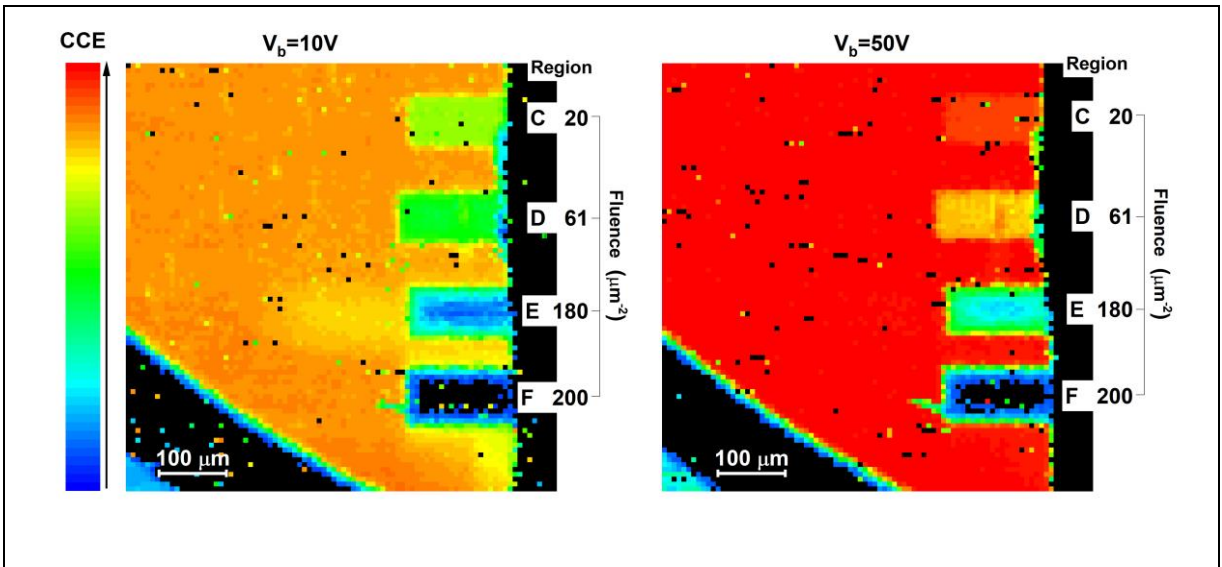

Fig. 5: frontal IBIC distributions using 20 MeV C ions taken with a bias voltage of 10 V and 50 after irradiation of 4 selected areas at different fluences

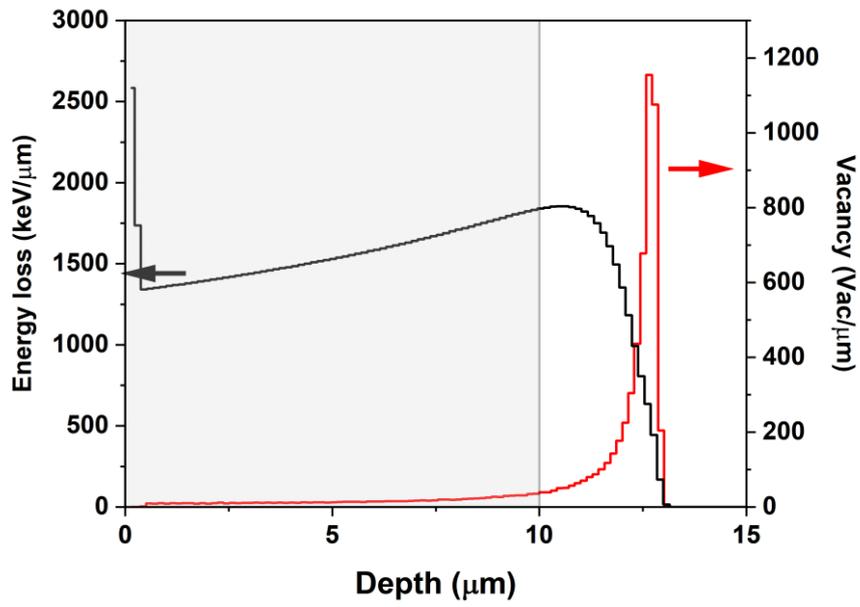

Fig. 6: Energy loss profile (left scale) and vacancy profile (right scale) of 20 MeV C ions in SiC, from SRIM simulation [30]. The grey rectangle indicate the extension of the depletion region @ 10 V bias voltage.

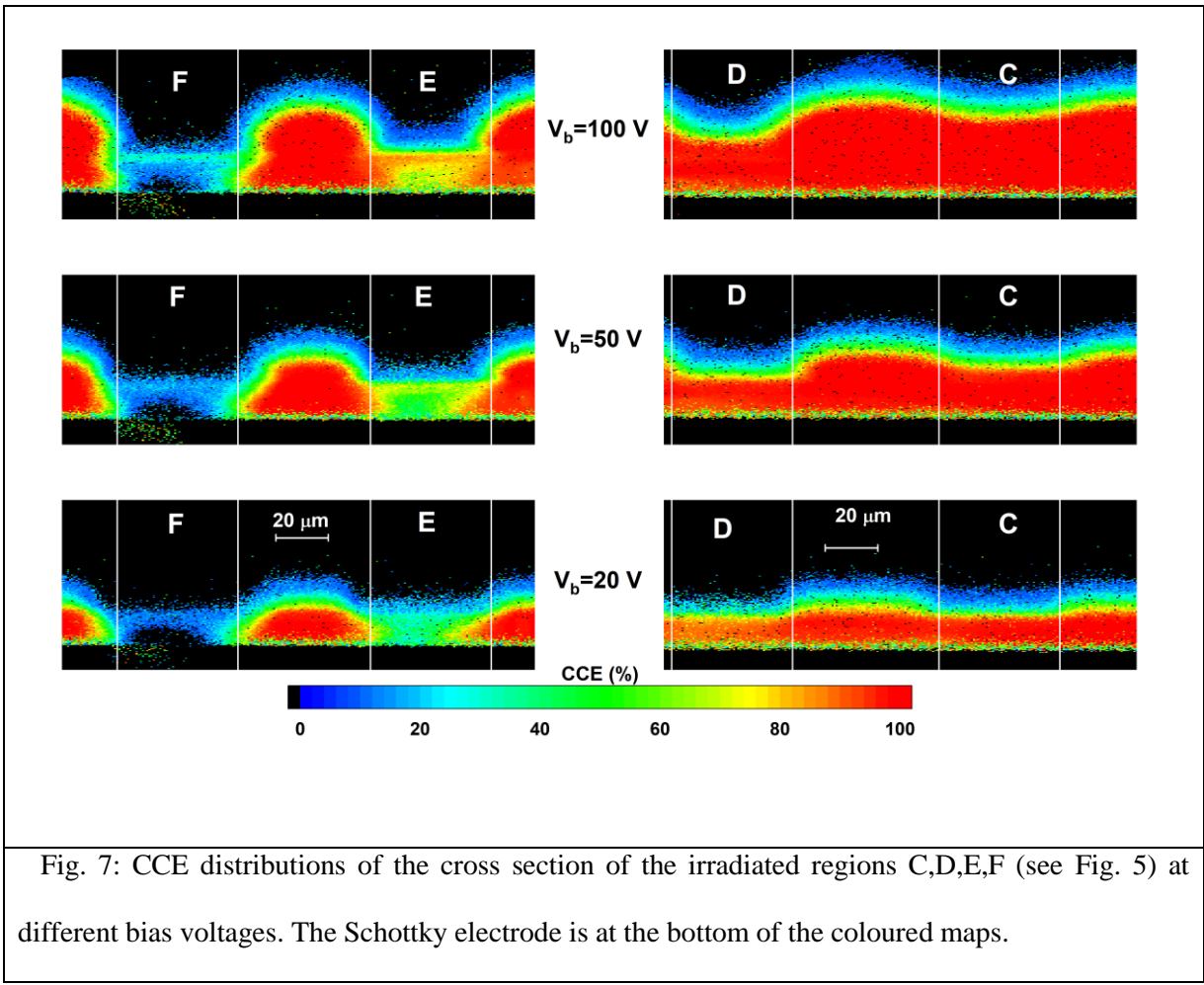

Fig. 7: CCE distributions of the cross section of the irradiated regions C,D,E,F (see Fig. 5) at different bias voltages. The Schottky electrode is at the bottom of the coloured maps.

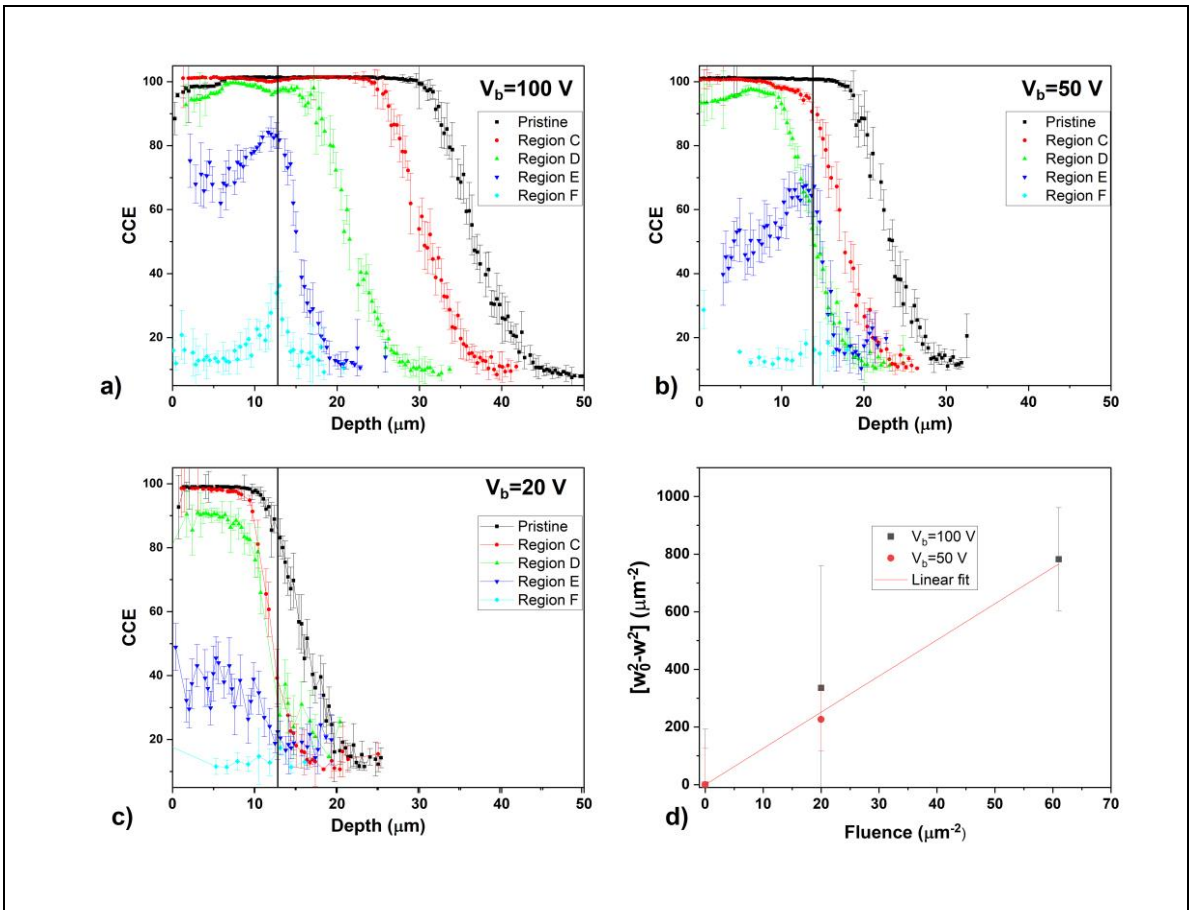

Fig. 8: CCE profiles extracted from the central part of the damaged regions shown in Fig. 7 for bias voltages $V_b$=100 V a), 50 V b), 20 V c). The vertical line at 12.8 μm indicates the position of the peak of the vacancy profile generated by 20 MeV C ions (see Fig. 6). d) difference in quadrature of the depletion layer of the C and D irradiated regions and the pristine region vs. the ion fluence. Black markers indicate values extracted from the CCE maps at $V_b$=100 V; red markers at $V_b$=50 V. The red line is the linear fit.